# Collective Excitations of Rotating Dipolar Fermi Gases in the Fractional Quantum Hall Regime

**Szu-Cheng Cheng**

**Department of Physics, Chinese Culture University, Taipei, Taiwan, R. O. C.**

We apply the magneto-roton theory of the fractional quantum Hall effect (FQHE) to study the collective excitation spectrum of rotating dipolar Fermi gases. The predicted spectrum has a finite energy gap in the long wavelength limit and a roton minimum at finite wave vector. There is a deepening of roton minimum in going from filling factor 1/3 to filling factor 1/5. Such a deepening of the minimum is a signature of incipient crystallization near filling factor 1/7. We also demonstrate that there are no low-lying single-particle excitations below the roton mode. The FQHE fluid of rotating dipolar fermions behaves as an incompressible superfluid at low temperature.

PACS number: 03.75.Ss, 73.43.-f, 73.43.Lp

The Fractional Quantum Hall Effect (FQHE) is a macroscopic quantum phenomenon occurring in a two-dimensional (2D) electron gas at high magnetic field [1, 2]. The ground state of the FQHE is a strongly correlated and incompressible liquid state (Laughlin's liquid state) [2], whose excitation spectrum has a large gap at the long wavelength limit and a deep magneto-roton minimum at finite wave vector [3]. The energy comparison between the Laughlin-liquid state and the solid (Wigner crystal) state indicated that the Wigner-crystal state is stable as the filling factor $\nu$ is below 1/7 [4, 5], where the filling factor $\nu$ is a number to measure the fraction that Landau levels are occupied by particles and defined by $\nu = 2\pi\rho a^2$, $\rho$ and $a$ being the average density and the magnetic length, respectively. This transition is also shown in the deepening of the magneto-roton minimum, which is a precursor to the gap collapse associated with Wigner crystallization near $\nu=1/7$ [3].

Recently there is a remarkable progress in the experiment to manipulate quantum many-body states in the ultra cold atomic gases. Systems with well understood and controllable interactions are available. The richer and more complex many-body states in the strong interaction regime are observable from the achievement involving Feshbach resonances [6] and the realization of a Mott insulator-superfluid transition of atoms in optical lattices [7]. It is also possible to realize strongly correlated states of the FQHE in a rotating Bose gas [8]. Rotating gases feel the Coriolis force in the rotating frame. The Coriolis force on rotating gases is identical to the Lorentz force of a charged particle in a magnetic field. Quantum-mechanically, energy levels of a charged particle in a uniform magnetic field are named as Landau levels and display discrete energy levels of a harmonic oscillator [9]. If interactions between neutral atoms are much less than the energy gap of Landau levels, the physical phenomena of rotating gases are within the lowest Landau-level (LLL) approximation. In the LLL approximation the only effect on the system is from interactions and the analogy of the FQHE for electrons is occurring in rotating gases.

Theoretically, we expect the existence of the FQHE states in the context of ultra cold atomic gases with short-range contact interactions [8]. In fact, it is difficult to realize the FQHE states experimentally, due to a small excitation gap with weak contact interactions. Since interactions between atoms localized in the lattices being strongly enhanced compared to the interaction of atoms in a trap, a novel method that uses atoms in optical lattice to raise the excitation gap of the system is proposed [10,11]. It is also possible to explore the potentially strong correlations induced by the long-range interaction. The recent realization of a concrete physical system plays a significant role in the creation of the dipolar interaction [12]. The dipolar interaction is a long-range interaction that puts dipolar gases correlated strongly [13]. It was argued that the Laughlin state for $\nu=1/3$ was incompressible with gapped excitations in rotating quasi-2D polarized dipolar gases [14]. There was a transition from the Laughlin liquid states to the Wigner crystal as the filling factor is below a critical filling factor [15].

In this letter we report the first evaluation of intra-Landau-level excitation energies of the FQHE states in rotating quasi-2D polarized dipolar gases whose mass is $M$. Dipolar gases are fully polarized along the rotating axial direction. The approach is similar to the single-mode approximation (SMA) used in calculations of intra-Landau-level excitation energies of the FQHE states in a 2D electron gas at high magnetic field [3]. We focus on low-energy excitations of the density oscillations in the extreme quantum limit where the ground state of a FQHE state is formed entirely within the LLL. To avoid the inhomogeneous effect in the 2D direction, we studied the system rotating in the limit of critical rotation, where the magnitude of the rotating frequency $\Omega$ of the system is close to but still smaller than the trapping frequency. Under the limit of critical rotation, the density of the trapped gas becomes uniform except at the boundary given by a trapping potential.

The formal development of mathematics within the subspace of the LLL has been presented elsewhere [16]. We shall apply previous formalisms and take the magnetic length $a = \sqrt{\hbar/2M\Omega} = 1$. The projected density operator with $N$ particles within the LLL is given by [3, 16]

$$\bar{\rho}_{\mathbf{k}} = \sum_{j=1}^{N} \exp\left[-ik\, \partial/\partial z_j\right] \exp\left[-ik^* z_j / 2\right], \qquad (1)$$

where $z_j = x_j + i y_j$ and $k = k_x + i k_y$ are the complex representations of the $j$th particle position and the wave vector of the density oscillations. The projected density operators satisfy a commutation relation defined by [3]

$$\left[\bar{\rho}_{\mathbf{k}},\, \bar{\rho}_{\mathbf{q}}\right] = \left(e^{k^* q/2} - e^{k q^*/2}\right) \bar{\rho}_{\mathbf{k}+\mathbf{q}}. \qquad (2)$$

We are now equipped with tools to project the Hamiltonian into the LLL from the definition of the projected density operator. The kinetic energy is neglected due to its irrelevant constant energy in the LLL approximation. The projected Hamiltonian is written as

$$H = \frac{1}{2(2\pi)^2} \int d^2\mathbf{q}\, V(\mathbf{q}) \left(\bar{\rho}_{-\mathbf{q}} \bar{\rho}_{\mathbf{q}} - \rho e^{-q^2/2}\right), \qquad (3)$$

where $V(\mathbf{q})$ is the Fourier transform of the quasi-2D dipolar-interaction potential with the coupling constant $D$ and expressed as [17]

$$V(\mathbf{q}) = \frac{D}{a^3}\left[\frac{4\sqrt{2\pi}}{3(z/a)} - 2\pi q W(\xi)\right], \qquad (4)$$

where $\xi = qz/\sqrt{2}a$, $z$ is the extension of dipolar gases in the axial direction, and $W(\xi) = \exp(\xi^2)\,\text{Erfc}(\xi)$ is related to the complementary error function $\text{Erfc}(\xi) = 1 - \left(2/\sqrt{\pi}\right)\int_0^{\xi} \exp(-t^2)\,dt$. We have a strictly 2D dipolar system when $z=0$ and $V(\mathbf{q}) = -2\pi q D/a^3$.

Following the previous SMA theory on the evaluation of excitation energies of the FQHE [3], we construct the collective-excitation state $|\mathbf{k}\rangle$ with momentum $\mathbf{k}$ as $|\mathbf{k}\rangle = N^{-1/2}\bar{\rho}_{\mathbf{k}}|0\rangle$, where $N$ is the particle number and $|0\rangle$ is the ground state with energy $E_0$. From $H|\mathbf{k}\rangle = E_k|\mathbf{k}\rangle$, we can evaluate the excitation energy $\Delta(k) = E_k - E_0$ by the formula $\Delta(k) = \bar{f}(k)/\bar{s}(k)$, where $\bar{f}(k)$ and $\bar{s}(k)$ are the projected oscillator strength and static structure factor, respectively. Using Eq. (3), $\bar{f}(k) = \langle 0|[\bar{\rho}_{-\mathbf{k}}, [H, \bar{\rho}_{\mathbf{k}}]]|0\rangle/2N$ is readily evaluated with the commutation relation given by Eq. (2) and we have

$$\bar{f}(k) = \frac{1}{2}\sum_{\mathbf{q}} V(\mathbf{q})(e^{q^*k/2} - e^{qk^*/2})$$
$$\times [\bar{s}(q)e^{-k^2/2}(e^{-qk^*/2} - e^{-q^*k/2})$$
$$+ \bar{s}(k+q)(e^{qk^*/2} - e^{q^*k/2})]. \qquad (5)$$

The projected static structure factor is related to the Fourier transform of the radial distribution for the ground state, $g(\mathbf{r})$: $\bar{s}(k) = \rho \int d^2\mathbf{r} \exp(-i\mathbf{k}\cdot\mathbf{r})g(\mathbf{r}) + \exp(-k^2/2)$. The radial function $g(\mathbf{r})$ of the Laughlin liquid state has been computed by the Monte Carlo method; it can be approximated as [3]

$$g(\mathbf{r}) = 1 - e^{-r^2/2} + \sum_{m=1}^{\infty} \frac{C_m}{m!}[1 - \exp(im\pi)]\left(r^2/4\right)^m e^{-r^2/4}, \qquad (6)$$

where the expansion coefficients $C_m$, obtained by fitting the Monte Carlo data, can be found in Ref. [3, 18]. Having obtained an analytic form of the radial function, ground-state energies and the projected static structure factors of the FQHE states are readily evaluated. The ground-state energies $E(\nu)$ for $\nu=1/3$ and $\nu=1/5$ are $E(1/3) = 0.3676(1/3)^{3/2} D/a^3$ and $E(1/5) = 0.3354(1/5)^{3/2} D/a^3$, respectively, as $z=0$. These energies are in excellent agreement with the energies from the Monte Carlo method [15].

From the Fourier transform of Eq. (5), we compute $\bar{s}(k)$, then use this to evaluate $\bar{f}(k)$ and calculate the collective-excitation energy $\Delta(k)$ finally. That $\bar{s}(k) \sim |\mathbf{k}|^4$ for small $k$ is a phenomenon of the lack of density fluctuations or the incompressibility of the FQHE states at long wavelength. This is the source of the finite energy gap observed in the FQHE of a 2D electron gas in a strong magnetic field. Therefore, we compute $\Delta(k)$ at small $k$ using the exact leading term in $\bar{s}(k)$, $\bar{s}(k) = (1-\nu)|\mathbf{k}|^4/(8\nu)$, in the long wavelength limit. We studied the collective-mode dispersion for different values of the dimensionless thickness parameter $z/a$ along the rotating axis. The evaluated $\Delta(k)$ versus wave vector $ka$ for $\nu=1/3$ and $\nu=1/5$ is shown in Fig. 1 and Fig. 2, respectively. The essential features of excitation dispersions exhibit a finite excitation-energy gap at $k=0$ and an energy minimum $\Delta$ (roton) at finite wave vector. A finite excitation-energy gap at $k=0$ implies that density fluctuations cost energy and a rotating-dipolar-Fermi gas is an incompressible fluid. The roton structure of excitations is caused by a peak in $\bar{s}(k)$, which is usually interpreted as a starting signal of crystallization [3]. There is a significant reduction in the roton energy from $\nu=1/3$ to $\nu=1/5$. The deepening of the minimum is a precursor of forming a Wigner crystal. This collapse of the gap to form a crystal state is shown by the roton minimum lying close to the primitive reciprocal-lattice wave vector $G$ of a Wigner crystal, which is given by $G_{1/3} = 1.56\ ka$ at $\nu=1/3$ and $G_{1/5} = 1.20\ ka$ at $\nu=1/5$, respectively. There is a phase boundary between the FQHE states and the Wigner-crystal states near $\nu=1/7$ [15]. Further evidence in favor of this interpretation of the roton minimum is provided by the fact that there is a significant reduction in the size of the gap as the experimental relevant thickness $d/a$ is increasing. The collapse of the gap happens at a critical thickness and the collapsed gap lies close to the primitive reciprocal-lattice wave vector of a Wigner crystal.

Having understood a physical picture of excitations, we now examine the validity of the SMA. The SMA theory, being based on a variational *ansatz*, provides an energy upper bound to the lowest excited state. From figures 1 and 2, the excitation energy at $k=0$ is larger than $2\Delta$. Excitation energies being higher than $2\Delta$ can easily break into two roton modes [19]. A two-roton bound state with total momentum $k=0$ is the lowest energy long-wavelength neutral excitation of the FQHE [3, 19]. The SMA is quantitatively accurate out to the roton minimum [3]. The SMA breaks down as many different states can couple to the density fluctuation for large wave vectors. Therefore we did not show excitation dispersions for wave vectors being much larger than the primitive reciprocal-lattice wave vector of a Wigner crystal. Though the SMA is invalid for large wave vector, it would be interesting to evaluate the first excitation moment $\Gamma$ in the limit of large $k$ [3]:

$$\Gamma = \frac{2[\nu E(1) - E(\nu)]}{1-\nu}, \qquad (7)$$

where $E(1)$ is the ground-state energy at $\nu=1$. For large $k$ the quasiparticle and the quasihole, which are a pair of a quasiparticle-quasihole excitation created by the density wave, are far apart. The upper bound of the excitation energy of a quasiparticle-quasihole pair is given by the first excitation moment $\Gamma$. Suppose the binding energy of a far apart quasiparticle-quasihole pair is very small due to the $1/r^3$ of the dipole-dipole interaction. The upper bound of the excitation energies of a quasiparticle plus a quasihole is also provided by the first excitation moment $\Gamma$. The results of the first excitation moment for different thicknesses of the system are shown in Fig. 3. $\Gamma$ is decreasing as the thickness of the system is increasing. The values of $\Gamma$ are above the roton gap, but lie considerably below the result of the quasiparticle excitation spectrum of Baranov et al. [14]: the excitation energy of a quasihole is $\Delta\varepsilon_{qh} = 0.9271\, D/a^3$ and the same order of magnitude is expected in the excitation energy of quasiparticles. Although we do not have an explanation of this large discrepancy, but we can certainly say that there are no low-lying excitations of quasiparticles and quasiholes below the roton mode. The lack of low-lying single-particle excitations means that there is no dissipation acting on the FQHE fluid of rotating dipolar fermions at low temperature. It is a superfluid at low temperature. Such phenomenon is a characteristic of bosonic superfluids: there are no single-particle excitations, and the dynamics of the system is completely described in terms of collective modes.

We studied collective excitations of rotating dipolar Fermi gases theoretically. The possible experimental ways to verify theoretical results will be the next important issue. One of the possible ways is using Bragg spectroscopy [20] to measure the excitation spectrum and the static structure factor of the FQHE. Bragg spectroscopy of a Bose-Einstein condensate in a trap has been realized experimentally [21]. The measured excitation spectrum agreed well with the Bogoliubov spectrum for a condensate. Hafezi *et al*. [11] also proposed a detection technique based on Bragg spectroscopy to obtain the excitation spectrum and the static structure factor of the FQHE in an optical lattice. Their study has shown that excitation gaps of the FQHE can be realized by Bragg spectroscopy.

In conclusion, we investigated the collective-excitation modes of a rotating quasi-2D dipolar system in the FQHE regime and estimated the upper bound of excitation energies of quasiparticle-quasihole pairs. The finite-thickness effects on excitations were considered. We demonstrated that the collective-excitation spectrum had a gap at the long wavelength limit and a roton minimum near the primitive-reciprocal-lattice wave vector of the Wigner crystal. The energy gaps of collective excitations are decreasing as the thickness of the system is increasing. The FQHE fluid of rotating dipolar fermions is incompressible due to a finite energy gap at $k$=0. The deepening of the roton gap in going from $\nu$=1/3 to $\nu$=1/5 is a signature of incipient crystallization near $\nu$=1/7. Excitation energies of quasiparticle-quasihole pairs were investigated by looking at the first excitation moment $\Gamma$. The values of $\Gamma$ are greater than the collective-excitation gaps. There are no low-lying excitations of quasiparticles and quasiholes in the FQHE states. The FQHE fluid behaves as a superfluid at low temperature.

The author acknowledges the financial support from the National Science Council (NSC) of Republic of China under Contract No. NSC96-2112-M-034-002-MY3. The author also acknowledges the support of the National Center for Theoretical Sciences of Taiwan during visiting the center.

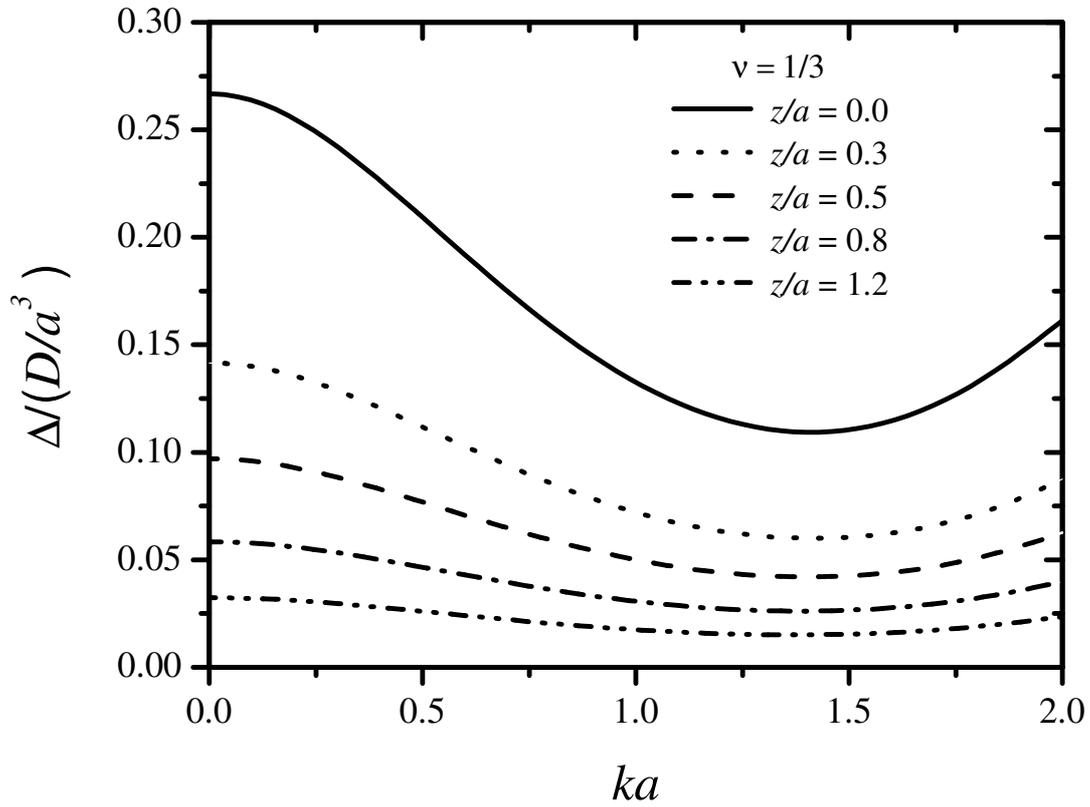

Fig. 1. Collective-excitation dispersions at $\nu=1/3$. The effect of finite thickness on collective-excitation dispersions is considered. Going from the top curve to the bottom, the values of the thickness parameter $z/a$ are 0.0, 0.3, 0.5, 0.8 and 1.2, respectively. The magnitude of primitive reciprocal-lattice wave vector of corresponding Wigner crystal is at 1.56 $ka$.

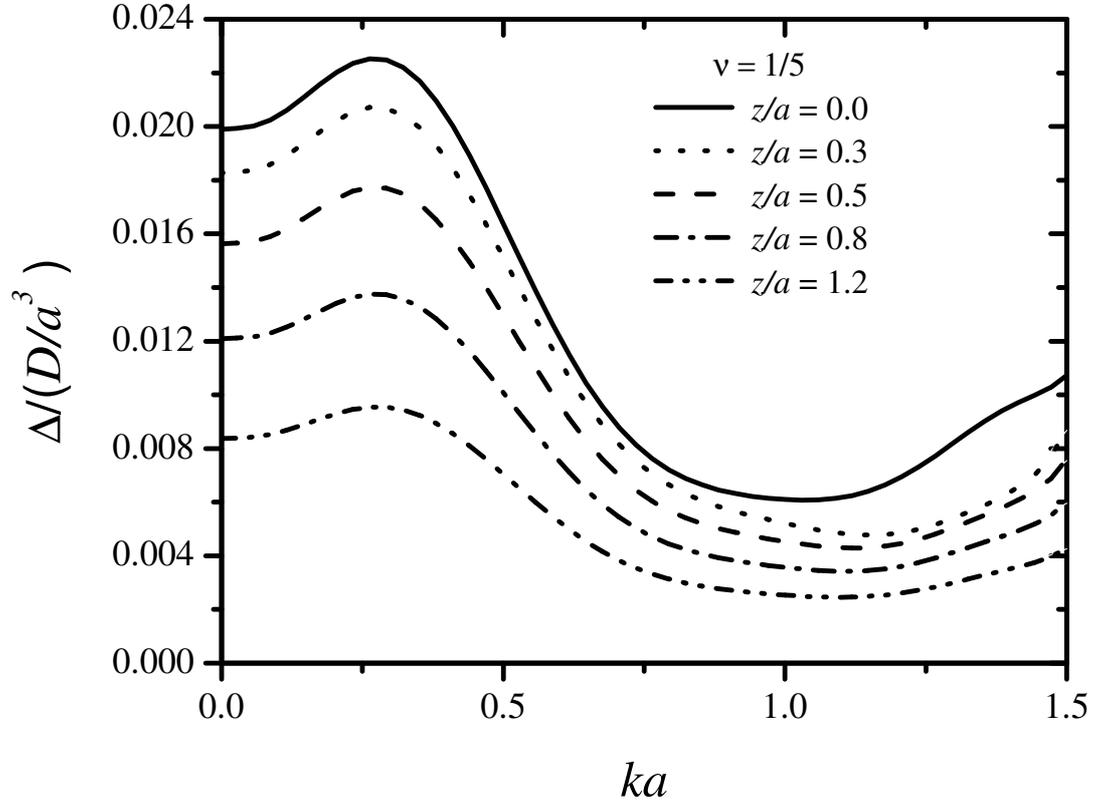

Fig. 2. Collective-excitation dispersions at ν=1/5. The effect of finite thickness on collective-excitation dispersions is considered. Going from the top curve to the bottom, the values of the thickness parameter $z/a$ are 0.0, 0.3, 0.5, 0.8 and 1.2, respectively. The magnitude of primitive reciprocal-lattice wave vector of corresponding Wigner crystal is at 1.20 $ka$.

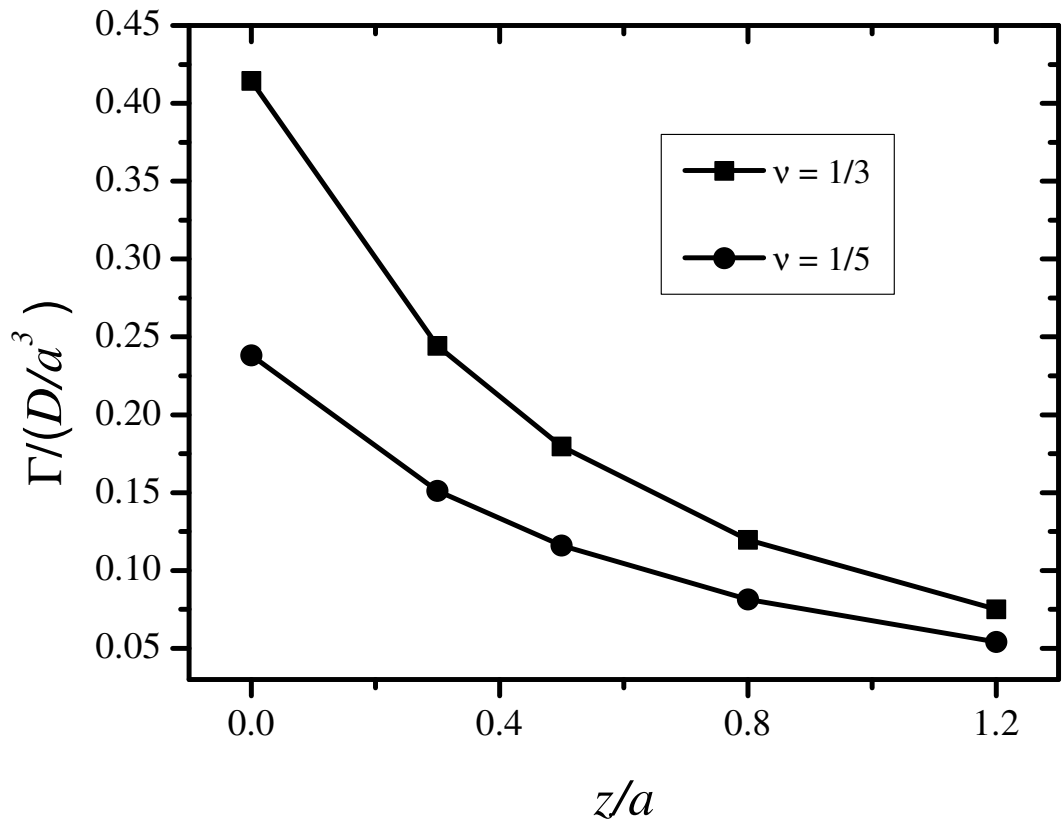

Fig. 3. First-excitation moment versus thickness parameter *z/a*. Squares and circles are the values of the first excitation moment at ν=1/3 and ν=1/5, respectively.